\documentclass[acmconf,authorversion]{acmart} 
\usepackage{booktabs}
\usepackage{graphicx}
\usepackage{multirow}
\usepackage[normalem]{ulem}
\useunder{\uline}{\ul}{}
\usepackage{rotating}
\usepackage{float}
\usepackage{inputenc}

\AtBeginDocument{%
  \providecommand\BibTeX{{%
    \normalfont B\kern-0.5em{\scshape i\kern-0.25em b}\kern-0.8em\TeX}}}

\copyrightyear{2021}
\acmYear{2021}
\setcopyright{acmlicensed}
\acmConference[CHI '21]{CHI Conference on Human Factors in Computing Systems}{May 8--13, 2021}{Yokohama, Japan}
\acmBooktitle{CHI Conference on Human Factors in Computing Systems (CHI '21), May 8--13, 2021, Yokohama, Japan}\acmDOI{10.1145/3411764.3445206} \acmISBN{978-1-4503-8096-6/21/05}

\begin{document}

\title[What Do We See in Them?]{What Do We See in Them? Identifying Dimensions of Partner Models for Speech Interfaces Using a Psycholexical Approach}

\author{Philip R. Doyle}
\affiliation{University College Dublin}
\email{philip.doyle1@ucdconnect.ie}
\author{Leigh Clark}
\affiliation{Swansea University}
\email{l.m.h.clark@swansea.ac.uk}
\author{Benjamin R. Cowan}
\affiliation{University College Dublin}
\email{benjamin.cowan@ucd.ie}

%%
%% By default, the full list of authors will be used in the page
%% headers. Often, this list is too long, and will overlap
%% other information printed in the page headers. This command allows
%% the author to define a more concise list
%% of authors' names for this purpose.
\renewcommand{\shortauthors}{Doyle, Clark \& Cowan}

%%
%% The abstract is a short summary of the work to be presented in the
%% article.
\begin{abstract}
Perceptions of system competence and communicative ability, termed partner models, play a significant role in speech interface interaction. Yet we do not know what the core dimensions of this concept are. Taking a psycholexical approach, our paper is the first to identify the key dimensions that define partner models in speech agent interaction. Through a repertory grid study (N=21), a review of key subjective questionnaires, an expert review of resulting word pairs and an online study of 356 users of speech interfaces, we identify three key dimensions that make up a users’ partner model: 1) perceptions towards partner competence and dependability; 2) assessment of human-likeness; and 3) a system’s perceived cognitive flexibility. We discuss the implications for partner modelling as a concept, emphasising the importance of salience and the dynamic nature of these perceptions.
\end{abstract}




\maketitle

\section{Introduction}
Through the growing use of devices like Amazon Echo and Google Home, speech agents have become common dialogue partners. Unlike embodied conversational agents (ECAs) or robots, speech agents rely heavily on voice as a primary form of interaction, lacking the embodiment required for common forms of non-linguistic communication (e.g. physical gestures) \cite{doyle_mapping_2019}. Speech agent interaction research has emphasised the importance of user’s perceptions toward a system’s competence and communicative ability as a dialogue partner (i.e. their partner models), impacting speech choices \cite{branigan_role_2011,cowan_whats_2019} and the types of tasks that users entrust speech agents with \cite{doyle_mapping_2019, luger_like_2016}. However, while the role of partner models is widely acknowledged \cite{branigan_role_2011, cowan_voice_2015, cowan_whats_2019,luger_like_2016, oviatt_linguistic_1998}, the concept is currently under-defined with regards to its underlying dimensions. 

Our paper contributes by being the first to define the key dimensions that constitute people's partner models for speech agents. Taking a psycholexical approach, our work gathered a set of word pairs to describe a person's partner model of speech agents, before using principal component analysis (PCA) to identify the dimensions that emerge from these word pairs. To achieve this we conducted two phases of item generation. In phase 1, we conducted a repertory grid study exploring perceptions of speech agents as dialogue partners among 21 users, providing 246 unique user-generated word pairs. In phase 2, we conducted a review of items from subjective questionnaires applicable to partner modelling related concepts. These included speech interface usability and user experience measures as well as socio-cognitive measures of concepts such as theory of mind and anthropomorphism, generating a further 155 word pairs. Following a screening process, 51 items were selected for use in an online questionnaire, used to measure speech agent perceptions. Through principal component analysis (PCA) of questionnaire responses from 356 users, we identify three key dimensions that form a user’s partner model in speech agent interaction: 1) Partner competence and dependability (emerging from perceptions of competence, reliability and precision); 2) human-likeness (whether the speech agent is perceived as human-like, warm, social or transactional); and 3) cognitive flexibility (whether the speech agent is perceived as flexible, interactive or spontaneous). For a full list of attributes within each dimension see Table 5. Our research is the first to outline and quantify the multi-dimensional nature of partner models for speech agent interaction. This constitutes a significant step in defining partner models as a concept, facilitating further elaboration of the theory, and providing a scaffold for future user-centered speech interface research.

\section{Related Work}
The following section presents a synthesis of theoretical accounts for partner modelling, research examining their role in speech interface interactions, and evidence for their impact on language production. Although our research is focused on speech agents, the work reviewed incorporates findings from robotics as well as findings from human-machine (HMD) and human-human dialogue (HHD).  

\subsection{The Construction and Dimensionality of Partner Models}
Rooted in research on perspective taking in HHD, partner models stem from the idea that people enter dialogue with assumptions about their interlocutors \cite{branigan_role_2011,clark_using_1996} and that these drive language choices in conversation \cite{brennan_two_2010, clark_using_1996, fussell_coordination_1992}. Conceptually, partner models might be thought of as mental models of a dialogue partner, yet there are differences in how these are conceptualised. Mental models are small-scale internal representations of the world and objects within it \cite{craik_nature_1943}. Whereas, partner models refer more specifically to a person's internal representation of an interlocutor's (human or machine) dialogic competence, considering their capabilities and knowledge as a "communicative and social being" \cite[p.~1]{cowan_what_2017}. Initially, these assumptions take the form of a broad \emph{global partner model}. This global model is triggered by a host of verbal and non-verbal cues, such as a speaker’s accent or language choices, age, gender and ethnicity \cite{branigan_role_2011, nickerson_how_1999}, and is initially based on broad stereotypes about the cultural groups an interlocutor is assumed to belong to \cite{branigan_role_2011,tobar-henriquez_lexical_2020}. Global models are then updated in accordance with direct experience, gradually leading to the construction of a more individualised \emph{local partner model} for a specific interlocutor \cite{brennan_two_2010}. 

Although partner models are seen as influential in HHD and HMD \cite{branigan_role_2011, cowan_whats_2019, cowan_they_2017, edlund_towards_2008}, studies in HMD tend to be relatively broad and unspecific when scoping the concept. Research has identified that users tend to see systems as at-risk listeners \cite{oviatt_linguistic_1998} or basic conversational partners \cite{branigan_role_2011} when compared to humans. Yet qualitative research suggests that, rather than being simplistic and unidimensional in nature, these models may be complex and multifaceted, constructed through attempts to understand both functional limitations and social relevance of speech technologies \cite{leahu_how_2013, doyle_mapping_2019}. This results in significant "...overlaps and blurrings between explanatory categories such as 'human' and 'machine'” \cite[p.~1]{leahu_how_2013}, with people's partner models in a constant state of flux as they attempt to rationalise their experiences with speech agents. This explanation is very similar to accounts of how global partner models are updated in the construction of more accurate local partner models \cite{brennan_two_2010}. It is also similar to socio-cognitive explanations of how mental models are updated, where two superordinate models are compared along relevant dimensions \cite{jacob_sociocognitive_1998,westbrook_mental_2006}.

Other research has noted how these models are heavily influenced by the human-likeness of speech systems. Superficial cues of human-likeness in speech agents, such as expressive synthetic voices \cite{akuzawa_expressive_2018} that use conversational rules and structures adapted from HHD \cite{edlund_pause_2009, gilmartin_social_2017}, prompt frequent comparisons with humans among users \cite{cowan_they_2017}. Indeed, human-like heuristic models seem to act as an anchor for users’ initial expectations \cite{cowan_they_2017,cowan_what_2017, doyle_mapping_2019}. For instance, in a study looking at perceived knowledge of landmarks, people's estimations of what humans and machines knew were strongly positively correlated, though people expected machine partners to have a wider breadth of knowledge \cite{cowan_they_2017}. Similar results have also been found in human-robot interaction (HRI) \cite{sau-lai_lee_human_2005}, whilst others have emphasised the importance of perceived anthropomorphism and intelligence in robot interactions \cite{bartneck_measurement_2009,salem_err_2013}. Collectively, this work suggests that the construction of a user's initial partner model may be significantly influenced by assumptions that are driven by the human-like design of speech agents, which sets high expectations for a system’s abilities and competence. However, the inaccuracy of this initial human-like global model is quickly identified by users, prompting comparisons that highlight the system’s inherent functional limitations. This gulf between a user’s initial expectations and their actual experiences creates cognitive conflict, leading to frustration, limited engagement \cite{luger_like_2016}, and subsequent updating of their partner model \cite{brennan_two_2010,leahu_how_2013,luger_like_2016}.

\subsection{The Importance of Partner Models for Interaction}
In addition to frustration caused by dissonance between people’s initial models and their actual experiences \cite{luger_like_2016}, there is ample evidence that partner models significantly influence language behaviour in HMD. This is commonly found in comparative studies of language in interactions with human and machine partners. When compared to HHD, in HMD people are shown to use more concise syntax \cite{amalberti_user_1993,bell_interaction_1999}, fewer anaphoric pronouns \cite{bell_interaction_1999} and less variation in dialogue strategies \cite{amalberti_user_1993,bell_interaction_1999}. Partner models have also been shown to influence a key linguistic phenomena known as lexical alignment \cite{tobar-henriquez_lexical_2020} - a tendency for dialogue partners to converge on the same lexical terms during dialogue. Specifically, people show stronger lexical alignment when they believe they are interacting with a computer compared to a human dialogue partner, and when they believe they are interacting with a basic computer compared to an advanced computer \cite{branigan_linguistic_2010,branigan_role_2011}. These results mirror earlier work showing stronger lexical alignment in interactions with basic systems \cite{pearson_adaptive_2006} and later work showing stronger alignment in interactions with avatar based virtual agents versus human dialogue partners \cite{bergmann_exploring_2015}. People have also demonstrated a higher likelihood of using American English terms to describe objects when interacting with an American accented speech system compared to an Irish accented speech system \cite{cowan_whats_2019}. Design cues used to signal and encourage anthropomorphism also influence changes in language behaviour. For instance, systems that use anthropomorphic dialogue strategies encourage increased levels of politeness, indirect phrasings and use of second person pronouns \cite{brennan_effects_1994}. These various forms of linguistic adaptation are thought to result from people using their partner models to hypothesize ways of ensuring communicative success, similar to the concept of audience design \cite{bell1984language}. 

\begin{figure*}[htbp]
    \centering
    \includegraphics[keepaspectratio, width=0.8\textwidth]{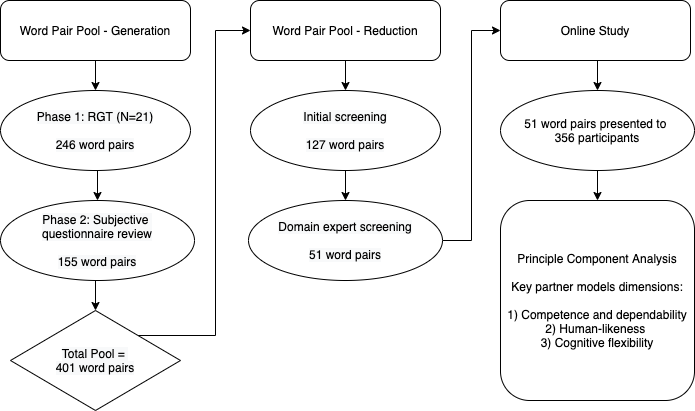}
    \caption{Overview of research approach and results}
    \Description{A flow chart outlining the research approach, beginning with word pair pool generation, ending with principle component analysis identifying the three key partner model dimensions. Details in supplementary materials.}
    \label{fig:Fig. 1}
\end{figure*}

\subsection{The Psycholexical Approach}
The psycholexical approach is the most well established and widely used method in psychology for identifying dimensions that underlie subjective constructs \cite{kline_handbook_2013}. With a long history in personality and individual differences research, the approach also underscores a number of different cognitive, psychoanalytic and behavioural techniques \cite{kline_handbook_2013}. Historically it has been used to distinguish the interpersonal traits of people and products, including technological artifacts \cite{hogan_blending_2013,volkel_developing_2020} and assistive technologies \cite{schaffalitzky_identifying_2009}. The basic tenet behind the psycholexical approach is that people's perceptions of an experience become encoded in their language \cite{volkel_developing_2020}, which can be accessed introspectively. This data can then be analysed using a variety of cluster analysis techniques to identify consistencies across the terms people use to define their perceptions, outlining the underlying dimensions of a given construct \cite{volkel_developing_2020}.

\subsection{Research Aims}
From the work discussed, it is clear that partner models play an important role in speech agent interaction. Yet current conceptualisations lack detail and dimensionality, which limits its utility as a concept. A more detailed explanation of partner models is crucial to future speech agent research. Further delineation of this concept is needed to help explain what drives speech interaction behaviours in more detail and elaborate on current accounts to better explain speech interaction phenomena \cite{clark_state_2019}. By uncovering the common salient dimensions of partner models, speech agent designers and researchers can potentially measure the impact of design changes on users' perceptions and behaviours. Through a multi-method psycholexical strategy, we aim to identify the key dimensions relevant to partner modelling in speech agent dialogue.

\section{Our Approach}
Following previous work \cite{volkel_developing_2020} we took a psycholexical approach to define and identify the dimensions of partner models, gathering a set of word pairs that describe the concept and then identifying clusters within these word pairs through PCA. Word pairs were generated over two phases. Phase 1 used the repertory grid technique (RGT) with 21 users generating word pairs relevant to their partner models, resulting in 246 unique word pairs. In phase 2 we added a further 155 word pairs based on a review of subjective questionnaire metrics used to measure partner modelling related concepts. Word pairs were then screened for duplicates as well as by two domain experts to identify the most relevant word pairs for conceptualising partner models. From this 51 items were retained and were given to 356 users to evaluate their experiences with speech agents through an online questionnaire study. We then used PCA to analyse questionnaire responses, identifying word pair clusters and the key partner model dimensions that emerge from these clusters. Further details of these stages and processes are outlined below, with an overall outline of the research shown in Figure \ref{fig:Fig. 1}.

\begin{figure*}[htbp]
    \centering
    \includegraphics[keepaspectratio, width=0.7\textwidth]{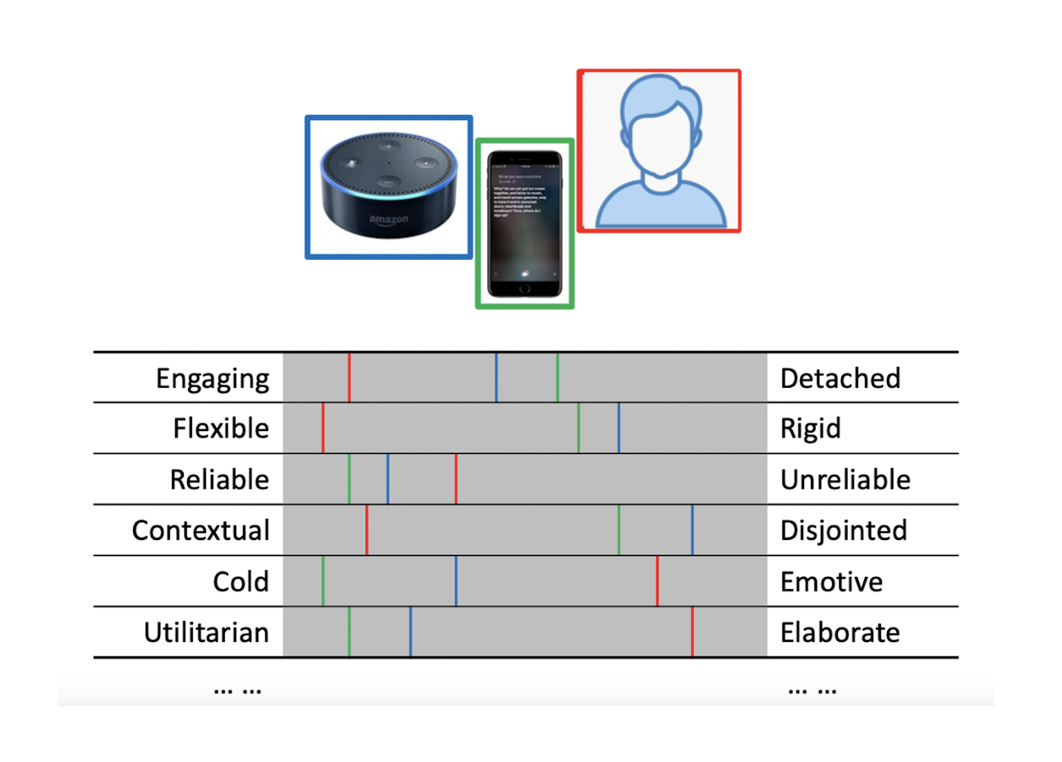}
    \caption{Example of repertory grid. Amazon Alexa, through Echo Dot (Blue), and Siri (Green) are compared to a human (Red) conversational partner. Coloured lines on repertory grid refer to ratings for each elements on constructs identified}
    \Description{A hypothetical example of a completed repertory grid. Implicit constructs are in the left hand column, emergent constructs are in the right hand column and lines representing where each dialogue partner sit between each of these poles.}
    \label{fig:Fig. 2}
\end{figure*}

\subsection{Word Pair Generation Phase 1: Repertory Grid}
\subsubsection{Research Design} Initial conceptualisation of partner models emphasises that they are perceptions of a dialogue partner’s communicative ability \cite{branigan_role_2011} and that, in the context of speech agent interaction, they appear to carry a strong initial anthropomorphic and social component \cite{cowan_they_2017, doyle_mapping_2019,leahu_how_2013}. To elicit items related to partner models we had people make direct comparisons between the communicative ability of speech agents and human interlocutors, which were gathered as part of a recent study \cite{doyle_mapping_2019}. The procedure involved use of the RGT \cite{kelly_psychology_1991}. Commonly used as part of personal construct theory in psychology \cite{kelly_psychology_1991}, RGT is an experience-orientated research approach designed to discover important latent dimensions of people’s perceptions towards particular people or objects \cite{kelly_psychology_1991, fransella_manual_2004, jankowicz_easy_2004}. Highlighted as a way to gather insight about how people conceptualise experiences, the technique requires participants to generate word pairs (termed personal constructs) that describe, conceptualise and compare particular objects of study (termed elements) \cite{kelly_psychology_1991}. When using the technique, participants are exposed to three elements at a time during a familiarization session, two similar and one dissimilar, through a paradigm known as triading. Triading is designed to make comparisons easier by making important characteristics more salient for participants \cite{fransella_manual_2004, jankowicz_easy_2004}. Construct elicitation comes next, where participants compile a list of words (a.k.a. implicit constructs) that best describe key similarities and differences between each of the elements, before identifying an appropriate opposite pole (a.k.a. emergent construct). This adds context for each implicit construct generated. During construct elicitation participants are asked to talk aloud, providing further context and reasoning around why they are choosing certain words and how it relates to their interactions. RGT therefore allows researchers an insight into an individual’s reasoning and conceptualising process for the elements presented in a study \cite{fransella_manual_2004}. The final phase is a rating phase, where participants rate where each of the elements sit between the various word pairs they provided. Historically, RGT has been used in educational psychology \cite{shaw_focus_1978} and information design \cite{hogan_blending_2013}. It has also been used to examine perceptions of website usability \cite{tung_attributes_2009}, strategic information systems \cite{cho_exploring_2010}, mobile technologies \cite{fallman_capturing_2010} and human-likeness in speech interfaces \cite{doyle_mapping_2019}. In HCI the technique provides a user-centered, exploratory approach that identifies how people define and describe their conceptualisation of technological artefacts \cite{fallman_capturing_2010}. This user-centered exploration was critical to ensure word pairs closely represented how users perceive speech interfaces as dialogue partners.

\subsubsection{Participants} 24 participants from a European university were recruited via email. Each was given a €10 honorarium for taking part. Three participants were omitted from the data due to difficulties completing the grids unassisted. Of the remaining 21 participants (f=9, m=11; mean age=23.1yrs, sd=5.49) all were native or near native English speakers. Relatively frequent speech interface users accounted for 38.1\% of participants (daily, a few times per week, or a few times per month), with people who use them rarely (38.1\%) or never (23.8\%) making up the rest of the sample. Among those that had used speech interfaces, Apple's Siri was most commonly used (50\%), followed by Google Assistant (31.3\%) and Amazon Alexa (18.8\%).

% Please add the following required packages to your document preamble:
% \usepackage{booktabs}
\begin{table*}[h]
\caption{Question types with examples}
\label{tab:Table 1}
\begin{tabular}{@{}cc@{}}
\toprule
\textbf{Question/request type} & \textbf{Question/request format} \\ \midrule
Conversational & \textit{\begin{tabular}[c]{@{}c@{}}How are you today?\\ Where are you from?\\ Tell me a joke\end{tabular}} \\ \midrule
Information retrieval & \textit{\begin{tabular}[c]{@{}c@{}}Who is {[}insert famous person's name{]}?\\ What is the square root of {[}insert three digit number{]}?\\ How do I get to the City Centre from here?\end{tabular}} \\ \midrule
Subjective/opinion-based & \textit{\begin{tabular}[c]{@{}c@{}}Do you like {[}insert favorite genre of music{]}?\\ Can you recommend a place to eat {[}insert favorite food when eating out{]}?\\ What do you think of {[}insert famous person's name - same as before{]}?\end{tabular}} \\ \bottomrule
\end{tabular}
\end{table*}

\subsubsection{Procedure} Upon arrival at the lab, participants were briefed about the nature of the study and what participation entailed, and were given details about their rights regarding participation and data protection. Next, they were asked to provide basic demographic information along with details about their speech interface usage. Then the familiarization phase began, where participants interacted with three different dialogue partners (elements): a human (a member of the research team) and two speech agents, namely Siri through a smartphone and Alexa using an Echo Dot smart speaker. The order of interactions with each dialogue partner was counterbalanced between participants, with interactions limited to a set of 9 predefined questions (see Table \ref{tab:Table 1}). Questions were designed to emphasise differences in the way these types of dialogue partners communicate; further prompting direct comparisons between the communicative capabilities of humans and speech agents. Following the familiarization phase participants were shown an empty grid and were asked to ‘write a list of words (implicit constructs) that best described the key similarities and differences between each of the dialogue partners (elements), focusing on their communicative abilities.’ If needing a further prompt participants were asked to generate words by focusing on ‘how you felt about the way each partner received and communicated information.’ In accordance with RGT protocol \cite{fransella_manual_2004}, the interviewer did not guide word generation, encouraging them to move on or return to a word if they were finding generation difficult. After compiling a list of implicit constructs, participants were then tasked with identifying a list of emergent constructs (i.e. an appropriate opposite word for each implicit construct). This lead to a word pair being created that the users feel reflects an important aspect of the communicative abilities of speech agents, relative to humans. Throughout this construct elicitation phase participants were encouraged to talk aloud, providing context and reasoning around why they were choosing certain words and how it related to their interactions. Finally there was a rating phase where participants placed each partner on a line between each of the word pairs. This was used to identify whether a particular implicit or emergent construct is more closely associated with human or machine dialogue partners, and provided context to support data analysis.

% Please add the following required packages to your document preamble:
% \usepackage{booktabs}
% \usepackage{graphicx}

\subsubsection{Results} Participants produced a total of 266 construct pairs, 246 of which were unique pairings. For brevity a sample of these word pairs are shown in Table \ref{tab:Table 2}, with a full list of word pairs available in supplementary materials.

%TABLE 1 HERE (REPGRID QUESTIONS)
% Please add the following required packages to your document preamble:
% \usepackage{booktabs}
\begin{table*}[h]
\centering
\caption{Sample word pairs generated from repertory grid study. Full word pair list are included in Supplementary Material}
\label{tab:Table 2}
\begin{tabular}{@{}c@{}}
\toprule
\begin{tabular}[c]{@{}c@{}}Opinionated/Non-judgmental; Biased/Neutral; Free/Bookish; Expansive/Limited; \\ Spontaneous/Pre-programmed; Colloquial/Universal knowledge; \\ Abstract/Specific knowledge; Lateral/Inflexible thinking; Personal relatability/Manufactured; \\ Genuineness/Ungenuine; Real/Fake; Canny/Uncanny; Emotional/Cold; \\ Personal/Robotic; Connection/Disconnected-disinterested; Engaged/Remote; \\ Humour/Humourless; Expansive/To-the-point; Convenience/Inconvenience; \\ Elaborate/Pointed; Polite/Blunt or rude; Colloquial/Formal; Vague/Detailed; \\ Two-way/One-way; Conversive/Monologue; Humanness/Machineness; Real/Organic-Artificial; \\ Personalised/Commercialised; No agenda/Agenda; To help/To serve\end{tabular} \\ \bottomrule
\end{tabular}
\end{table*}

\subsection{Word Pair Pool Generation Phase 2: Subjective Questionnaire Review}
\subsubsection{Research Design and Procedure} Findings from the RGT study provide a strong starting point, with 246 word pairs produced. However, to ensure the set of word pairs provided comprehensive coverage, we also conducted a review of relevant subjective questionnaires. This involved a review of all subjective questionnaire metrics identified in a recent systematic review of speech interface research in HCI \cite{clark_state_2019}. We also conducted a Google Scholar search for subjective questionnaires used to measure concepts related to partner modelling, namely: theory of mind (ToM); mental models; perspective taking; metacognition; anthropomorphism and dehumanisation; and social-cognition. Each of these topics was used as a search term, prefaced by the terms ‘questionnaire’, ‘survey’ and ‘subjective measure’. After reviewing a total of 75 measures, 44 were identified as containing items that could contribute to the pool of word pairs being generated. These included established and bespoke HCI usability measures used in previous speech and HMD research (n=17), and established measures from socio-cognitive psychology covering the range of topics outlined above (n=27). Contributing questionnaires and specific items co-opted from them are included in Table \ref{tab:Table 3}. A full list of questionnaires and co-opted items are provided in supplementary materials. The vast majority of the measures reviewed here adopted Likert scale response options, many in conjunction with semantic differential scales similar to what participants produced using RGT in Phase 1. Where questionnaire items were in the form of a short phrase (e.g. “The system was pleasant” - SASSI \cite{hone_towards_2000}), the key adjective 'pleasant' was extracted and an appropriate antonym was generated either from other items on the same scale or by the lead author using a thesaurus.

\subsubsection{Results} The review yielded a further 155 word pairs: 86 word pairs coming from speech interface and HMD usability and user experience metrics, and 67 from established measures in socio-cognitive psychology. When combined with the RGT results, the word pair pool after both generation phases stands at a total of 401 pairs of words, which were then screened as outlined below.

\begin{sidewaystable*}
\vspace*{+19cm}
% Please add the following required packages to your document preamble:
% \usepackage{booktabs}
% \usepackage{graphicx}
\centering
\caption{Subjective Questionnaire Review: HCI measures used in previous speech and HMD research (denoted as **), and established measures from socio-cognitive psychology (denoted as *)}
\label{tab:Table 3}
\resizebox{\textwidth}{!}{%
\begin{tabular}{@{}ccc@{}}
\toprule
\textbf{Research area and measures reviewed} &
  \textbf{Sample of measures reviewed} &
  \textbf{Sample of extracted words} \\ \midrule
 
 \begin{tabular}[c]{@{}c@{}} **Subjective questionnaires used in previous HMD research \\ Potential items taken from 5 \cite{hassenzahl2003attrakdiff, hone_towards_2000, polkosky_toward_2005,polkosky2003expanding, bos1999survey} \\ out of 8 \cite{parasuraman1991refinement, brooke1996sus, schrepp2017design} reviewed \end{tabular} &
  \begin{tabular}[c]{@{}c@{}}AttrakDiff2: Product appeal based on hedonic \\ and pragmatic qualities \cite{hassenzahl2003attrakdiff}; \\ SASSI: Usability and Interactivity \cite{hone_towards_2000}; \\ MOS-X: Quality of synthetic speech \cite{polkosky_toward_2005}; \\ SUISQ: User Satisfaction \cite{polkosky2003expanding} \end{tabular} &
  \begin{tabular}[c]{@{}c@{}}people-centric/technical, practical/impractical, cumbersome/facile, \\ unmanageable/manageable, isolates/connects, stylish/lacking Style, \\ poor quality/high quality, excludes/draws you in, ugly/pretty\\ appealing/unappealing, consistent, efficient, repetitive, \\ lose track, harsh, raspy, strained \end{tabular} \\ \midrule
  
  \begin{tabular}[c]{@{}c@{}} **Bespoke metrics used in previous HMD research \\ Potential items taken from 11 \cite{larsen_assessment_2003,casali_effects_1990, dintruff1985user, walker_what_1998, lee_designing_2003, gong_shall_2001,dulude_automated_2002, qvarfordt_role_2003, clark_multimodal_2016} \\ \cite{cowan_what_2017, dahlback_similarity_2007, evans_impact_2010} out of 12 \cite{walker1998evaluating} reviewed \end{tabular} &
  \begin{tabular}[c]{@{}c@{}} Usability \cite{walker1998evaluating}; \\ User Acceptance of SDS \cite{casali_effects_1990, dintruff1985user}; \\ System Preference \cite{gong_shall_2001, walker_what_1998}; \\ User satisfaction \cite{dulude_automated_2002}; \\ User experience \cite{clark_multimodal_2016, cowan_what_2017}. \\\end{tabular} &
  \begin{tabular}[c]{@{}c@{}} understandable/incomprehensible, accessible/inaccessible, \\ appropriate/inappropriate, familiar/unfamiliar, \\ predictable/unpredictable, basic/advanced, capable/incapable, cheap/expensive, \\ good/bad, inflexible/flexible, lacks power/powerful, quick/slow, \\ stable/unstable, amateurish/professional, modern/old fashioned, \\ efficient/inefficient, trustworthy/untrustworthy,\end{tabular} \\ \midrule 
\begin{tabular}[c]{@{}c@{}} *Theory of Mind \\ Potential items taken from 12 \cite{lalonde_false_1995,baron-cohen_friendship_2003,wheelwright2006predicting,zoll2010questionnaire,spreng_toronto_2009,diaz2013amsterdam,roberts1991development} \\ \cite{goodman_psychometric_2001,sarason_assessing_1983,jette1986functional,grootaert_measuring_2004} out of 24 measures \cite{kinderman1996new,peterson1982attributional,brune2005emotion, rieffe2010assessing, slaughter2002theory,ajzen2006constructing} \\ \cite{keaton2017interpersonal,bass2004multifactor,gresham1990social,broadbent1982cognitive,jorm1994short,steger2006meaning} reviewed\end{tabular} &
  \begin{tabular}[c]{@{}c@{}} EQ \& TEQ: Assessing emotional empathy \cite{zoll2010questionnaire,spreng_toronto_2009};\\ FQ: Assessing attitudes toward friendships \cite{baron-cohen_friendship_2003}; \\ SQ-R: Assesses demand for understanding underlying rules \cite{wheelwright2006predicting};\\ ARSQ: Self-report of resting-state cognition \cite{diaz2013amsterdam}; \\ SSQ: A measure of social support \cite{sarason_assessing_1983}; \\ SC-IQ: Measure of social capital \cite{grootaert_measuring_2004}.\end{tabular} &
  \begin{tabular}[c]{@{}c@{}}interrupting, monopolizes conversation, apologetic \\ charitable/humility, meticulous, curious, lonely, upset \\ irritated/irritable, tired, sleepy, friend, confidant \\ superior, distractible, persistent, reflective, helps out \\ caring, lies, good friend, popular, solitary \\ dependable, be yourself around, appreciates you \\ console you, nervous, peaceful, downhearted, blue, \\ social with others, similar goals, source of expertise/advice, \\ willing, able, critical \end{tabular} \\ \midrule
\begin{tabular}[c]{@{}c@{}} *Mental Models and Partner Models \\ Potential items taken from 5 \cite{pejtersen2010second,PERRY2013587,MARTIN200348,knobloch_relational_2005,collins1996working} \\ out of 8 \cite{doi:10.1111/bjop.12050,BROCKMYER2009624, smeyer_mentalStates} measures reviewed\end{tabular} &
  \begin{tabular}[c]{@{}c@{}} MTQ48: Self-report measure of mental toughness \cite{PERRY2013587}; \\ Psi-Q: Self-report measure of mentalizing behaviours \cite{doi:10.1111/bjop.12050};\\ HSQ: Self-report measure of humor style \cite{MARTIN200348};\\ AAS-r: Attitudes toward romantic relationships \cite{collins1996working}; \\ RUQ: Perceived relational uncertainty \cite{knobloch_relational_2005}. \end{tabular} &
  \begin{tabular}[c]{@{}c@{}}supportive, burdensome, influential, \\ shows initiative, nervous, persistent, beautiful, \\ offensive, enthralling, self-deprecating, \\ criticizing, impressive, amusing, \\ absurd, reluctant, committed\end{tabular} \\ \midrule
  \begin{tabular}[c]{@{}c@{}} *Metacognition \\ Potential items taken from 1 \cite{cartwright-hatton_beliefs_1997} out of 5 \\ \cite{vandergrift2006metacognitive,KLEITMAN2007161,Fernie_meta_beliefs, Pintrich_learn,schraw1994assessing} measures reviewed\end{tabular} &
  \begin{tabular}[c]{@{}c@{}} MCQ: Beliefs about worries and intrusive thoughts \cite{cartwright-hatton_beliefs_1997}; \\ MSLQ: Motivational orientation for cognitive \\ motivation for engagement and self-regulation \cite{Pintrich_learn};\\ MAI: Awareness of metacognitive processes \cite{schraw1994assessing}. \end{tabular} & 
  \begin{tabular}[c]{@{}c@{}} glib, arrogant, selfish, complacent \\ copes, assertive, misleading, \\ confident, embarrassing\end{tabular} \\ \midrule
\begin{tabular}[c]{@{}c@{}} *Anthropomorphism and Dehumanisation\\ Potential items taken from 7 \cite{bartneck_measurement_2009,waytz_who_2010,ruijten2019perceived,crowne1960new,neave_influence_2015,salem_err_2013, forster2017increasing} \\ out of 12 \cite{chin2005developing, chin2004measuring, bruder2013measuring, barrett1996conceptualizing, darwin2011belief} measures reviewed \end{tabular} &
  \begin{tabular}[c]{@{}c@{}} Godspeed V: Anthropomorphism in HRI \cite{bartneck_measurement_2009};\\ IDAQ, AAS \& ATS: Tendencies to self-engage \\ in anthropomorphic behaviors \cite{waytz_who_2010,chin2004measuring,chin2005developing}; \\ CMQ \& CTQ: General tendency to \\ believe in conspiracies \cite{bruder2013measuring,darwin2011belief}; \\ MCSDC: Attempts to present oneself in \\ an overly positive manner \cite{crowne1960new}.\end{tabular} &
  \begin{tabular}[c]{@{}c@{}}life-like, elegant, inert, apathetic, awful \\ agitated, lethargic, intentional, free, conscious, \\ organized, purposeful, self-conscious, ambitious, imaginative, \\ qualified, intense, doubtful, courteous, comforting \\ alarming, trustworthy, deceptive, merciful, \\ deliberate, sentient, curious, fun-loving, sociable \\ aggressive, impatient, jealous, humble\end{tabular} \\ \midrule
  \begin{tabular}[c]{@{}c@{}} *Perspective taking and social cognition\\ Potential items taken from 2 \cite{olderbak2017emotion, buhrmester_five_1988} \\ out of 3 \cite{reniers2011qcae} measures reviewed\end{tabular} &
  \begin{tabular}[c]{@{}c@{}} QCAE: Self-reported cognitive and affective empathy \cite{reniers2011qcae}; \\ ESEQ: Assesses emotion specific affective and cognitive empathy \cite{olderbak2017emotion}; \\ ICQ: Self-reported interpersonal competence \cite{buhrmester_five_1988}.\end{tabular} & 
  \begin{tabular}[c]{@{}c@{}} infectious, acquaintance, unreasonable \\ embarrassing, confrontational, companion\end{tabular} \\ \bottomrule
\end{tabular}%
}
\end{sidewaystable*}

\subsection{Word Pair Pool Screening}
\subsubsection{Procedure}
Initial screening was carried out by the lead author to remove duplicates and near duplicates (word pairs that offered little semantic differentiation; e.g. ‘simple/complex’ and ‘simple/complicated’- only ‘simple/complex’ was retained). Word pairs that were considered too esoteric or vague were also adjusted (e.g. unfettered/restricted became free/restricted) or removed (e.g. conjunctive/uncoordinated). Multiple word expressions (N=15) were simplified (e.g. ‘Responsive or adaptive/Rigid or fixed response’. to ‘Adaptive/Fixed’). Transcriptions of talk aloud data were used to ensure accurate transformation. Finally, word pairs that were deemed obviously unrelated to the concept of partner modelling (e.g. infectious/uncommunicable), were removed in accordance with best practice guidelines on item pool screening \cite{kline_handbook_2013}. These ensure that the word pairs retained provide adequate coverage, retaining as much nuance between them as possible, whilst ensuring retained pairs are relevant to the concept being addressed. This initial screening process reduced the pool of items from 401 to 127 word pairs. This drop is largely accounted for by a high degree of redundancy when both item pools were combined. In cases where word pairs were similar to the RGT generated pairs, the user-generated RGT pairs were prioritised.

The remaining 127 word pairs were then systematically screened independently by two researchers with expertise in HCI, speech interaction, partner modelling, dialogue and socio-linguistics research. To guide the screening process researchers were provided with Kline's \cite{kline_handbook_2013} guidelines (outlined above) and a working definition of partner modelling (outlined below). The working definition was derived from a literature review of seminal work on mental models (e.g. \cite{craik_nature_1943, johnson-laird_mental_2010,johnson-laird_mental_1980, norman_design_2013, norman_observations_1983}), early work examining partner models in HHD and HMD interactions (e.g. \cite{branigan_role_2011,brennan_two_2010,cowan_understanding_2014,cowan_they_2017,duran_toward_2016}), and definitions of ToM (e.g. \cite{baron-cohen_friendship_2003}). The definition is designed to capture the dynamic \cite{brennan_two_2010, johnson-laird_mental_2010}, adaptive \cite{brennan_two_2010, johnson-laird_mental_2010}  and multidimensional \cite{doyle_mapping_2019,johnson-laird_mental_2010} nature of partner modelling, with a focus on perceptions of functional, cognitive and empathetic qualities of a dialogue partner that, according to ToM literature \cite{baron-cohen_friendship_2003}, are likely to influence interactions. It also incorporates key influences on partner models found in dialogue research, namely: stereotypes about the cultural communities a dialogue partner might belong to, and direct experience interacting with a particular dialogue partner \cite{branigan_role_2011,clark_using_1996}. Both are regarded as fundamental sources of information in formulating and updating global and local partner models, respectively \cite{brennan_two_2010}.\\

\begin{quote}
\emph{The term partner model refers to an interlocutor’s cognitive representation of beliefs about their dialogue partner’s communicative ability. These perceptions are multidimensional and include judgements about cognitive, empathetic and/or functional capabilities of a dialogue partner. Initially informed by previous experience, assumptions and stereotypes, partner models are dynamically updated based on a dialogue partner’s behaviour and/or events during dialogue.}\\
\end{quote}

Along with the definition, and best practice guidelines \cite{kline_handbook_2013}, the researchers were also provided with a spreadsheet containing the remaining 127 word pairs (see supplementary material). Researchers were instructed to review the pool independently, indicating which word pairs they felt were relevant, not relevant and items they were unsure about. In cases where they were unsure they were asked to comment on their reason for being unsure, providing details as to whether they were unsure about one or both terms in a word pair, and/or why they felt it was not suitable (i.e. not relevant to the concept, too vague or esoteric, or a more appropriate item is already contained within the pool).

% Please add the following required packages to your document preamble:
% \usepackage{booktabs}
% \usepackage{graphicx}
\begin{table*}[h]
\centering
\caption{Retained word pairs after screening phase. Full list of retained and eliminated items are included in Supplementary Material.}
\label{tab:Table 4}
\resizebox{\textwidth}{!}{%
\begin{tabular}{@{}c@{}}
\toprule
\begin{tabular}[c]{@{}c@{}}Authentic/Fake; Emotional/Clinical; Concise/Verbose; Subjective/Objective;  \\ Expert/Amateur; Empathetic/Apathetic; Reliable/Uncertain; Illogical/Logical; \\ Authoritative/Unsure; Flexible/Inflexible; Dependable/Unreliable; Assertive/Submissive; \\ Colloquial/Formal; Warm/Cold; Efficient/Inefficient; Human-like/Machine-like; Interactive/Start-stop; \\ Life-like/Tool-like; Adaptive/Fixed; Precise/Vague; Contextual/Non-contextual; Competent/Incompetent; \\ Personal/Generic; Hesitant/Decisive; Two-way/One-way; Assistant/Servant;  Intelligent/Unintelligent; \\ Elaborative/To-the-point; Misleading/Honest; Repetitive/Versatile; Meandering/Direct; Restricted/Free; \\ Abstract/Concrete; Basic/Advanced; Capable/Incapable; Sincere/Insincere; Consistent/Inconsistent; \\ Social/Transactional; Trustworthy/Untrustworthy; Confident/Uncertain; Spontaneous/Predetermined;\\ Cooperative/Uncooperative; Ambiguous/Clear; Broad/Specific; High Feedback/Low Feedback; \\ Predictable/Unpredictable; Amusing/Serious; Engaged/Disinterested; Complex/Straightforward; \\ Free/Restricted; Repetitive/Versatile; Authentic/Fake; Feedback High/Feedback Low\end{tabular} \\ \bottomrule
\end{tabular}%
}
\end{table*}

\subsubsection{Results}
The two domain experts independently agreed upon the retention of 24 word pairs and the rejection of 26 word pairs. The experts then met, along with the lead author, to discuss areas of disagreement (87 word pairs). Following the discussion a further 27 word pairs were retained leaving a total of 51 to be included in an online questionnaire. Table \ref{tab:Table 4} shows all retained items following the screening process. All eliminated items are included in the supplementary material.

\subsection{Quantifying Perceptions: Online Study and Principal Component Analysis}

\subsubsection{Research Design}
The next step involved presenting the 51 word pairs to participants through an online survey, which they used to rate their past experiences with the speech agent they interacted with most frequently. Word pairs were presented in the form of a questionnaire. Taking this empirical approach allows for the identification of word pair clusters. These then dictate the underlying structure of the concept with the strongest common terms in each cluster determining the meaning/context of a given dimension. Given the nature of the data produced using RGT, and that most measures reviewed used a similar response structure, we opted to use a 7-point semantic differential scale. Like with the RGT, this creates a scale were participants indicate where they feel speech agents sit between two opposite word poles.

\subsubsection{Participants}
390 participants completed the online questionnaire, recruited through email, posters and social media. Participants who completed the questionnaire were entered into a €200 voucher prize draw. From the 390, 34 participants were excluded due to heavily patterned responses that lack variation (i.e. more than 70\% of the same response option, or 90\% across just 3 response options) which is seen as evidence of inattentiveness \cite{MANIACI201461}. This means that 356 participants were included in the final analysis. All participants (f=61.5\%, m=36.8\%, non-binary or prefer not to say=1.7\%; age range= 18-70yrs, mean age= 28.5yrs, sd= 10.9) were required to have strong English reading and comprehension proficiency. Within the sample, 35.4\% had completed graduate or post-graduate education, 32.3\% had completed an undergraduate degree and 29.5\% had completed secondary and/or vocational education (remainder preferred not to say).

Participants reported moderate levels of experience with speech interfaces (7 point Likert scale; 1=very infrequent to 7=very frequent; mean=3.8, sd=2.06), using 2.6 (sd= 1.3) different devices on average to access them. Speech agents were by far the most common type of speech interface used, with Apple’s Siri being the most frequently accessed (80.1\%, [N=285]), followed by Google Assistant (64.3\%), Amazon Alexa (58.4\%) and Microsoft's Cortana (20.8\%). Use of multiple speech interfaces was common, with most participants having used two (39\%) or three (23.9\%) different interfaces, 29.2\% having used only one and 7.9\% (N=28) having used four or five. Accessing speech agents across multiple devices was also quite common (mean=2.6, sd=1.3), with 26.9\% of participants accessing them using between 3 to 6 different devices. Our sample most commonly accessed speech interfaces through smartphones (88.5\%) or smart speakers (59.2\%), followed by telephony based speech systems (30.6\%), laptops (28.1\%), in-car assistants (25.8\%) and tablets (22.2\%).

\subsubsection{Procedure}
\begin{figure*}[htbp]
    \centering
    \includegraphics[keepaspectratio, width=0.8\textwidth]{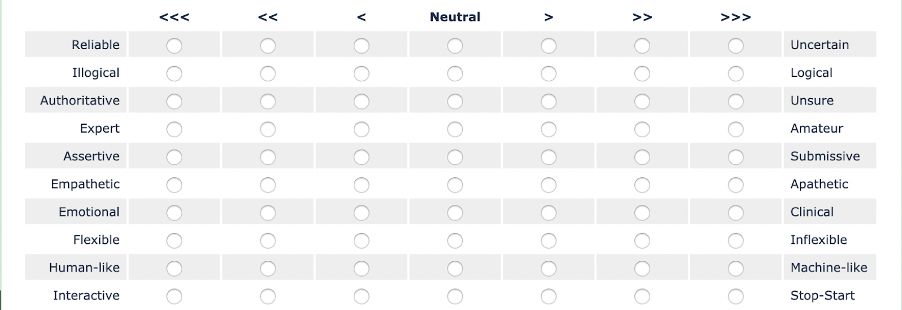}
    \caption{Example questionnaire structure}
    \Description{An example of the questionnaire structure as shown to participants. One word from a pair is in the left hand column. It's counterpart is in the right hand column, with a seven point scale in between.}
    \label{fig:Fig. 3}
\end{figure*}

The questionnaire was presented to participants online, using LimeSurvey. After following the link provided in recruitment materials, participants were presented with an information sheet giving full details of the study and their rights in relation to participation and data protection. After giving explicit consent to participate, participants completed a demographic questionnaire gathering information about their age, sex, educational attainment, nationality and their experience with speech interfaces. They were then presented with the 51 word pairs, each separated by a 7 point scale (see Figure \ref{fig:Fig. 3}). The display of word pairings was pseudo-randomised. Reflecting on previous interactions with speech agents, participants were asked to think about the way speech agents communicate with them and then rate the communicative ability of the speech agent they used most frequently on a scale between each of the word pairs displayed. Instructions were given to read each pair of words carefully, to respond as quickly and accurately as possible, and to try and avoid giving too many neutral responses. Participants were then fully debriefed as to the nature and aims of the study. 

\begin{table*}[]
\centering
\caption{Factor loadings for 3 factor PCA. Only loadings above 0.3 are displayed. All items removed during PCA are included in Supplementary Material.}
\label{tab:Table 5}
\resizebox{\textwidth}{!}{%
\begin{tabular}{@{}cccc@{}}
\toprule
\textbf{Items} &
  \textbf{\begin{tabular}[c]{@{}c@{}}Factor 1: \\ Partner Competence\\ \& Dependability\end{tabular}} &
  \textbf{\begin{tabular}[c]{@{}c@{}}Factor 2:\\ Human-likeness\end{tabular}} &
  \textbf{\begin{tabular}[c]{@{}c@{}}Factor 3:\\ Cognitive Flexibility\end{tabular}} \\ \midrule
Competent/Incompetent                                                              & 0.76  &      &       \\
Dependable/Unreliable                                                              & 0.68  &      &       \\
Capable/Incapable                                                                  & 0.68  &      &       \\
Consistent/Inconsistent                                                            & -0.67 &      &       \\
Reliable/Uncertain                                                                 & 0.66  &      &       \\
Ambiguous/Clear                                                                    & 0.65  &      &       \\
Meandering/Direct                                                                  & -0.64 &      &       \\
Expert/Amateur                                                                     & -0.64 &      &       \\
Efficient/Inefficient                                                              & 0.64  &      & 0.32  \\
Misleading/Honest                                                                  & 0.63  &      & -0.34 \\
Precise/Vague                                                                      & -0.62 &      &       \\
Cooperative/Uncooperative                                                          & 0.54  &      &       \\
Human-like/Machine-like                                                            &       & 0.75 &       \\
Life-like/Tool-like                                                                &       & 0.75 &       \\
Warm/Cold                                                                          &       & 0.68 &       \\
Empathetic/Apathetic                                                               &       & 0.66 &       \\
Personal/Generic                                                                   &       & 0.62 &       \\
Authentic/Fake                                                                     &       & 0.56 &       \\
Social/Transactional                                                               &       & 0.54 &       \\
Flexible/Inflexible                                                                &       &      & 0.66  \\
Interactive/Start-stop                                                             &       &      & 0.61  \\
Interpretive/Literal                                                               &       &      & 0.56  \\
Spontaneous/Predetermined                                                          &       &      & 0.51  \\ \midrule
\textbf{Eigenvalues}                                                               & 5.45  & 3.57 & 2.18  \\
\textbf{Proportion Variance}                                                       & 24\%  & 16\% & 9\%  \\
\textbf{Cumulative Variance}                                                       & 24\%  & 39\% & 49\%  \\
%\textbf{Proportion Explained}                                                     & 0.49  & 0.32 & 0.19  \\
%\textbf{Cumulative Proportion}                                                    & 0.49  & 0.81 & 1.00  \\ \bottomrule

\hline
\multicolumn{1}{c}{\textit{Factor Correlations}} &
   &
   &
   \\

\textbf{\begin{tabular}[c]{@{}c@{}}Factor 1: Partner Competence \& Dependability\end{tabular}} &
  - &
  0.21 &
  0.11 \\
\textbf{\begin{tabular}[c]{@{}c@{}}Factor 2: Human-likeness\end{tabular}}        & 0.21  & - & 0.36  \\
\textbf{\begin{tabular}[c]{@{}c@{}}Factor 3: Cognitive Flexibility\end{tabular}} & 0.11  & 0.36 & -  \\ \bottomrule
\end{tabular}}
\end{table*}

\subsubsection{Data Analysis}
We conducted a Principal Component Analysis (PCA) using the \textit{psych} \cite{Revelle_psych_2020} and \textit{GPArotation} \cite{Bernaards_GPArot_2005} packages in R (Version 1.1.456) \cite{R} so as to identify the dimensions present in the 51 word pairs. The primary purpose of PCA is to reduce the dimensionality of multivariate data, allowing for a large number of variables to be summarized within smaller subsets, or factors \cite{bryant_principal-components_1995,cowan_measuring_2014}. PCA was deemed most suitable as it does not require an \textit{a priori} hypothesized or predetermined factor structure, making it ideal for exploratory analysis \cite{bryant_principal-components_1995,cowan_measuring_2014}. We note that various recommendations are made regarding what constitutes a suitable sample size for conducting reliable PCA. A minimum sample size of 100 is required \cite{kline_psychometrics_2000}, with little difference seen in resultant factor structures when samples exceed 200 participants \cite{field_discovering_2013}. Based on this our sample of 356 is deemed suitable for PCA. 

Based on best practice guidelines to ensure reliable and clear factor structures \cite{clark2016constructing}, we first removed word pairs with weak inter-item correlations before conducting the analysis. Using established thresholds \cite{clark2016constructing}, word pairs with low mean inter-item correlations (r < .15) were removed, resulting in 14 word pairs being eliminated and 37 word pairs being included in the PCA. Kaiser-Meyer-Olkin (KMO) test of sampling adequacy was high overall (KMO= .91) for the remaining data, and across word pairs (KMO range= .95 to .81). Bartlett's test was also statistically significant [x2(666) = 4913.29, p<.001.] suggesting the data was suitable for PCA analysis.

Following \cite{field_discovering_2013}, a first PCA iteration was conducted with all items (word pairs) set as factors, to produce eigenvalues that are used to assess the number of factors to be retained. Here, the number factors retained in the rotated PCA was based on parallel analysis using the Hornpa \cite{hornpa_2015} function in R. Considered a more robust approach than traditional methods such as scree plots or Kaiser criterion \cite{field_discovering_2013}, in parallel analysis the number of factors that have higher eigenvalues than a set of simulated eigenvalues are retained. Simulated eigenvalues are generated from the original data set, with the number of simulations being set as a parameter (here 1000 simulations were run) \cite{field_discovering_2013}. Results of the parallel analysis suggested that 3 factors should be retained.

Next, PCA was conducted setting the number of factors to 3 and using direct oblimin rotation, the approach recommended when underlying dimensions are likely to be related \cite{field_discovering_2013}. Based on best practice guidelines \cite{field_discovering_2013, clark2016constructing} we then iteratively removed word pairs with weak communalities (<0.4), weak loadings (<0.5) and multiple cross loadings \cite{clark2016constructing}, until close to a mean communality of 0.5 was achieved. This led to a further 14 word pairs being removed. Following pruning of these 14 pairs the 3 factor model exhibited acceptable fit (0.96), acceptable mean item complexity (1.3), acceptable squared residuals (0.05) and accounted for 49\% of variance within the data. The final word pair clusters and factor structure are shown in Table \ref{tab:Table 5}. The factors revealed by the 3 factor model reflect dimensions that describe perceptions of: partner competence and dependability; partner’s human-likeness; and partner’s cognitive flexibility. Details regarding word pairs eliminated during PCA are included in the supplementary material.

\section{Discussion}
Our research took a psycholexical approach in mapping partner models as a concept, identifying key dimensions that constitute a user’s partner model for speech agents. First, through using the repertory grid technique (RGT), a total of 246 unique word pairs were generated by users to describe their partner models of speech agents. This data was complemented by a further 155 word pairs identified through a search of subjective questionnaires applicable to partner modelling related concepts. After screening the 401 word pairs, a selection of 51 word pairs were included in an online study of 356 speech agent users. These users were asked to rate the ability of speech agents as dialogue partners based on previous experience. Through principal component analysis (PCA), where a further 28 word pairs were eliminated, three key dimensions of a user’s partner model for speech agents were identified. These key dimensions reflected perceptions of a dialogue partner's: 1) competence and dependability (emerging from perceptions of competence, reliability and precision); 2) human-likeness (whether the speech agent is perceived as human-like, warm, social or transactional); and 3) cognitive flexibility (whether the speech agent is perceived as flexible, interactive or spontaneous). For a full list of attributes within each dimension see Table 5. This is a significant contribution in that it not only outlines the multidimensional nature of partner models in speech agent interaction, but adds specific structure to the concept that, to-date, has been lacking.

\subsection{The Influence of Design on Partner Models}
Our study adds much needed definition to the concept of user partner models. This should allow researchers to gather deeper insight into how design decisions may influence these models. Earlier work hypothesises that design choices, such as accent \cite{cowan_whats_2019} and anthropomorphic dialogue strategies \cite{brennan_effects_1994} affect partner modelling. Yet to date, it has not been possible to identify what specific aspects of a partner model are influenced by these choices, with studies using behavioural adaptation as evidence of general model change and influence \cite{cowan_whats_2019, branigan_role_2011}. Our work opens the possibility that these design decisions do not universally impact a user's model, being more nuanced in their effect. For instance, rather than influencing cognitive flexibility judgements, accent-based design choices may alter estimates of partner knowledge (relevant to competence and dependability) and human-likeness, making those dimensions more likely drivers of linguistic adaptation proposed \cite{cowan_whats_2019}. Echoing recent work, human-likeness in design tends to inform initial partner model development \cite{cowan_whats_2019,cowan_what_2017, doyle_mapping_2019}. To ensure partner models are accurate, human-like design should be congruent with the level of system capability \cite{moore2017spoken}. Our work gives a framework to help identify how human-like design choices may impact perceptions of human-likeness alongside other associated partner model dimensions such as perceptions of cognitive flexibility, and competence and dependability. The dimensionality identified is also useful for informing how other design choices may influence partner models. For example, expressive synthesis \cite{buchanan2018adding} and the use of more social talk \cite{gilmartin_social_2017} are likely to have an influence on specific model dimensions. Our work is an important first step in allowing researchers to explore how specific design choices affect these models more specifically. It is important to note that, rather than suggesting designers implement these partner model dimensions in speech interfaces, our findings identify perceptions that may be influenced by design changes.

\subsection{Partner Model Dimensionality, Salience and Dynamics}
Our findings emphasise that people’s partner models are clearly more detailed and complex than more general descriptions of speech agents as at risk listeners \cite{branigan_linguistic_2010,cowan_voice_2015,cowan_whats_2019,luger_like_2016,oviatt_linguistic_1998}, poor \cite{cowan_understanding_2014} or basic dialogue partners \cite{branigan_role_2011}. While the number of dimensions that are reflected on simultaneously is open to debate \cite{johnson-laird_mental_2010, norman_observations_1983}, it is likely that dimensions may becoming more or less salient in different contexts and over the course of interaction. For instance, the salience of dimensions may vary within certain situational contexts, such as when using an agent in health, wellbeing or care domains \cite{kocaballi_2020, spillane_2019} where human-likeness and perceptions of empathy are important. Indeed, events during speech agent dialogue may also bring dimensions to the fore, such as negotiating errors and miscommunications highlighting capability and flexibility judgements. The idea that certain aspects of a partner model will be more or less salient in response to specific system behaviours, dialogue events or contexts is similar to the idea of one-bit processing \cite{brennan_two_2010}. It also echoes accounts of how inaccurate mental models are amended \cite{johnson-laird_mental_2010}, and how partner specific information is incorporated in perspective taking \cite{duran_toward_2016}. All suggest that, models of a partner (or object) need not be comprehensive at all times, with specific dimensions dominating perceptions at different moments during the interaction or in response to dialogue events. With our research now identifying dimensions of speech agent partner models, future work can build on this by examining the influence of specific interaction events and context on model use. It also opens avenues for exploring how partner models might impact language production dynamically during HMD.

\subsection{The Interdependence and Dynamism of Partner Models}
Although this work significantly expands on the dimensionality of partner models as a concept, our results do not make any inferences about the causal relationships between the dimensions identified. However, it is highly likely that, although distinct, these dimensions are interdependent, with changes in one dimension impacting or affecting changes in another. For instance, it may be that changes to the perceived human-likeness of a system may lead to increases in perceptions of partner competence and dependability. This is eluded to in recent research, whereby the human-likeness of systems is seen to act as an anchor for initial perceptions of what a system knows and can do \cite{cowan_what_2017, luger_like_2016}. Work suggests that early attention to anthropomorphic characteristics leads to high expectations in regard to competence and dependability, which are quickly identified as unrealistic following interactions \cite{luger_like_2016}. Whilst work examining dynamic adaptations of partner models in response to dialogue events has been somewhat limited to-date, available accounts support our assertion that partner models are adaptive. For example, Leahu et al. \cite{leahu_how_2013} suggest that people use broad partner types (e.g. human and machine) to make comparisons across specific dimensions (i.e. humor and/or intelligence), whilst dynamically working towards a more accurate model \cite{leahu_how_2013}. Human dialogue work \cite{brennan_two_2010} also emphasises that partner models may evolve as a user's initial stereotype driven perceptions (e.g. global model) are fashioned into a more accurate, experience-based local model specific to a dialogue partner. Similar effects may occur within speech agent dialogue, where a user's initial perception of an agent becomes more nuanced once informed by direct experiences with a particular agent over time. An open question also relates to how these more nuanced models may then feedback to influence a user's global model to inform initial interactions with new, unfamiliar speech agents. Findings from our work open avenues for examining the interdependent and dynamic relationship between partner model dimensions, with a level of detail that was not previously possible. Future research efforts should focus on exploring how perceptions on these dimensions change over time, how they become more nuanced with experience and how this experience may feedback to inform global models of speech agents.

\section{Limitations and future work}
The triading paradigm used in RGT requires participants to be presented with three exemplars, two similar and one dissimilar, to provoke reflection about key characteristics of an object of interest (speech agents) and how they may be similar or different to an appropriate comparator (a human). Although users readily make comparisons between humans and machines in speech agent interaction without being instructed to do so \cite{luger_like_2016,cowan_they_2017}, the triading may have made these more likely. Previous work emphasizes that human comparison is core to speech agent partner model building and research \cite{branigan_role_2011,leahu_how_2013,luger_like_2016,cowan_voice_2015,cowan_what_2017}. Following this, we used a human comparator to prompt word-pair generation in word pool generation phase 1. In the later online study, responses to these word pairs were given specifically in relation to speech agent interaction only. Future work could add more speech agent elements - such as other speech interfaces and/or social robots with more or less human-like qualities - to gather a wider range of constructs, adding further granularity. 

To ensure initial word pairs accurately reflected speech agent perceptions, participants in the RGT study supplied words after direct interactions. For the online questionnaire study participants were asked to reflect on past experiences, rather than an interaction experienced directly prior to responding. This reflective approach was deemed most appropriate for building a general account of partner models as it reduces the potential for the online questionnaire responses being influenced by a specific agent or interaction encounter. 

Through the execution of the study we produced a set of 401 word pairs that describe a user’s partner model of speech agents. Much like in personality research, where the psycholexical approach is commonly used, the items produced are not only helpful in categorising and understanding the dimensionality of partner models, but can also form the basis of a self-report metric for measuring them. The current study is a significant step in developing such a questionnaire as it produces the item set and gives us an initial potential factor structure. Our future work aims to further develop the final word pair set into a fully validated partner modelling questionnaire. To do this we aim to conduct work to assess scale reliability (e.g. internal consistency and test-retest reliability) and validity (concurrent, discriminant and predictive validity testing), whilst performing confirmatory factor analysis on future datasets to ensure that the factor structure identified in this paper is robust \cite{kline_psychometrics_2000}. High factor loading items could be used as building blocks for a short-form scale, although this would need to be statistically validated.

Whilst statistical approaches like PCA can result in the loss of some rich qualitative insights, their aim is to ensure robust clustering of word pairs to identify emergent factors. Further work could add to our dataset, through research with additional speech agents, to identify additional dimensions.

Although our work has relevance for robotics and virtual agent research, it is also important to note that our scope is limited to identifying partner model dimensions for non-embodied speech agents, where speech is the primary if not exclusive form of communication. Work examining perceptions of embodied agents highlights unique considerations that may be incorporated in partner models when interacting with robots \cite{bartneck_measurement_2009,salem_err_2013} or avatars \cite{balas2018measuring}, such as animacy and/or safety. These are underpinned by the embodied nature of these interaction paradigms. Further work should look to replicate and build on our work within these domains. 

\section{Conclusion}
As the ubiquity of speech interfaces continues to increase, more people are now engaging with speech agents on a daily basis. Although research has consistently emphasised the importance of our perceptions toward a system’s capability as a dialogue partner (i.e. our partner models) in guiding interaction, the concept as it currently stands is poorly defined. Our work aimed to give structure to this concept by identifying the key dimensions of a user’s partner model. Through principal component analysis we identified that partner models for speech agents hold three key dimensions, which focus on perceptions of a dialogue partner's competence and dependability, human-likeness and apparent cognitive flexibility. This not only adds granularity, clarity and definition to the concept, but also highlights that there are multiple dimensions for designers to consider when aiming to support users and improve their interaction experience. 

\section{Acknowledgements}
This research was supported by an employment based PhD scholarship funded by the Irish Research Council and Voysis Ltd (R17830).

\bibliographystyle{ACM-Reference-Format}
\bibliography{CHI2021_9853ID}

\end{document}